\documentclass[preprint,12pt]{elsarticle}

\usepackage{amssymb}
\usepackage{epsfig}
\usepackage{graphicx}
\usepackage{bm}

\usepackage{color}
\usepackage{xcolor}
\usepackage{ulem}

\newcommand{\eqref}[1]{(\ref{#1})}
\newcommand{\Tr}{{\rm Tr}}
\newcommand{\Or}{{\rm O}}

\journal{Physica B}

\begin{document}

\begin{frontmatter}

\title{Phase separation in strongly correlated electron systems with wide and narrow bands: a comparison of the Hubbard-I and DMFT approximations}

\author{A O Sboychakov\corref{cor1}}
\address{Institute for Theoretical and Applied Electrodynamics,
Russian Academy of Sciences, 125412 Moscow, Russia}

\cortext[cor1]{E-mail: sboycha@mail.ru}


\begin{abstract}
The spinless Falicov-Kimball model on the simple cubic lattice is analyzed in the Hubbard-I and dynamical mean field (DMFT) approximations. The Matsubara and real frequency itinerant electron Green's functions, the evolution of the system with doping, and the range of phase separation are found in two approximations. At large values of the on-site Coulomb repulsion both approximations give similar results. The phase separation can be also favorable for a more general model, where heavy electrons have a finite bandwidth. This indicates that the phase separation phenomenon is an inherent feature of the systems described by  the Hubbard-like models with wide and narrow bands.
\end{abstract}

\end{frontmatter}

\section{Introduction}

Strongly correlated electron systems, the systems where the interaction between electrons is larger than their kinetic energy, are intensively studied nowadays. Heavy fermion compounds~\cite{ColemanHF}, high $T_c$ cuprates~\cite{CuRev}, magnetic oxides such as manganites~\cite{Dagotto}, cobaltites, and many others, belong to this class of materials. In theoretical study, the Hubbard model and its modifications such as $t-J$ model and others are widely used to explain the complex behavior of strongly correlated materials. The main problem arising in the analysis of the Hubbard-like Hamiltonians is the absence of reliable analytical methods at large electron-electron interaction. The numerical studies such as Monte-Carlo simulations and exact diagonalization on small clusters have large computational costs. On the other hand, there exists a huge number of different approximate methods not requiring time-consuming calculations. However, each of these methods has a restricted range of applicability, and it is not always clear whether the results obtained by this or that approximation reveal an actual physics or just reflect the features of initial assumptions.

The simple example of the models of correlated electrons is the Falicov-Kimball model. It was proposed in seminal paper of L.\,M.~Falicov and J.\,C.~Kimball, \cite{FalKim}, to describe the metal to insulator transition in rare-earth and transition-metal compounds. It includes both itinerant and localized electrons with on-site interaction between them. This model exhibits many interesting features such as non Fermi-liquid behavior~\cite{NonFL} for any value of coupling constant $U$, charge-density wave phase~\cite{Brandt,GruberCDW,FreericksCDW_PS}, and phase separation~\cite{FreericksCDW_PS,Maska,FreericksPS,WePRL}. The Falicov-Kimball model has been studied both numerically with Monte-Carlo simulations and exact diagonalization~\cite{Maska,Lemanskii,deVries}, and by using different approximations~\cite{Brandt,Kennedy,Plischke}. Several exact results have been obtained also, such as the existence of the phase separation at large coupling constant~\cite{FreericksPS}.

The phase separation is a quite ubiquitous phenomenon in the strongly correlated electron systems. It manifests itself in different situations (both in experiment and theory), where the itinerancy of charge carriers competes with their tendency to the localization. However, revealing the phase separation in a given system or model is not a trivial task, which usually demands a lot of numerical and analytical work. The first most reasonable step is the using of the analytical mean-field type approximations, which do not require laborious computations. If we reveal the tendency to the phase separation in the framework of these approximations, a problem arises, whether this result is an artefact of the approximation, or it captures the main physics and remains qualitatively unchanged in a more refined (but much more cumbersome) approximations such as dynamical mean-field theory (DMFT) or in numerical simulations.

To address this problem, in this paper I use as a testing ground the Falicov-Kimball model for which the DMFT results are quite transparent and compare DMFT with the Hubbard-I approximation. More precisely, I analyse the Hubbard-I and DMFT approximations of the spinless Falicov-Kimball model in three dimensions in the strong $U$ limit. The Hubbard-I approximation~\cite{hubbard1} first proposed  for the Hubbard model can be applied to the Falikov-Kimball model as well~\cite{WePRL}. This is a simple mean-filed-like scheme, which allows to obtain analytical results. Although this approximation fails for $U\lesssim W$, where $W$ is the bandwidth of itinerant electrons, it can be considered as a good approximation at large $U$. A more advanced approach, dynamical mean-field theory~\cite{DMFTreview}, has been  applied first to the Falicov-Kimball model in~\cite{Brandt}, and has been considered later in many theoretical papers (for a review, see, e.g.,~\cite{FKreview}). This approximation takes into account the local correlations more rigourously, but at the same time it requires more computation costs. Here I show that for strong coupling, $U\gg W$, both approximations give similar results for the band structure of itinerant electrons, the evolution of the system with doping, and the region of the phase separation.

After this, I analyze the possibility of the phase separation for a more general model, where `localized' electrons have a finite bandwidth $W_f$. This model can be considered as a two band Hubbard model for spinless fermions. The itineracy of `localized' electrons makes DMFT calculations much more cumbersome in comparison to that for the Falicov-Kimball model. But again, both approximations give similar results for the range of the phase separation. It is shown that the phase separation for this model exists when $W_f<W_f^{*}$, where $W_f^{*}\sim0.3W$, although the Hubbard-I and DMFT approximations give slightly different values for the $W_f^{*}$. This correlates well with the result of~\cite{TwoBand}, where the phase separation in the spin-$1/2$ two-band Hubbard model is shown in the Hubbard-I approximation. So, the results presented in this paper indicate that the phase separation phenomenon is the inherent feature of multiband Hubbard models with wide and narrow bands, and not an artefact of this or that approximation scheme.

\section{Falicov-Kimball model in the Hubbard-I approximation}\label{SecHI}

The Hamiltonian of the Falicov-Kimball (FK) model is~\cite{FalKim}
\begin{eqnarray}\label{HFK}
H&=&-\sum_{ij}\left(t_{ij}d^{\dag}_{i}d_{j}+H.c.\right)-\epsilon_{f}\sum_{i}n^{f}_{i}+U\sum_{i}n^{d}_{i}n^{f}_{i}\nonumber\\%
&&-\mu\sum_{i}(n^{d}_{i}+n^{f}_{i}),
\end{eqnarray}
where $n^{d}_{i}=d^{\dag}_{i}d_{i}$, $n^{f}_{i}=f^{\dag}_{i}f_{i}$, and $d^{\dag}_{i}$ ($d_{i}$) and $f^{\dag}_{i}$ ($f_{i}$) are creation (annihilation) spinless fermion operators for the band and localized electrons, respectively. The first term in~\eqref{HFK} describes the kinetic energy of the band electrons, the second term controls the position of the localized level with respect to the center of the conduction band, while the third term corresponds to the on-site interaction between the band and localized electrons. In further analysis I will consider attractive interaction with $U>0$.

The Hubbard-I approximation is based on the decoupling in the system of equations for the electrons Green's functions. The use of Matsubara Green's functions is more convenient to study the system at finite temperatures. It also allows simply comparing the Hubbard-I and DMFT approximations, since DMFT deals with Matsubara one-particle Green's functions. The equation of motion for one particle $d$-electron Green's function $G_d(\tau-\tau',\,i-j)=-\langle\hat{T}d_i(\tau)d^{\dag}_{j}(\tau')\rangle$ is:
\begin{eqnarray}\label{Gd}
&&\left(-\frac{\partial}{\partial\tau}+\mu\right)G_d(\tau,\,i-j)+\sum_{l}t_{il}G_d(\tau,\,l-j)-\nonumber\\
&&U\,F_d(\tau,\,i-j)=\delta_{ij}\,\delta(\tau)\,,
\end{eqnarray}
where $F_d(\tau-\tau',\,i-j)=-\langle\hat{T}n^f_{i}(\tau)d_i(\tau)d^{\dag}_{j}(\tau')\rangle$ is the two-particle Green's function. The equation of motion for $F_d$ is the following:
\begin{eqnarray}\label{Fd}
&&\left(-\frac{\partial}{\partial\tau}+\mu-U\right)F_d(\tau,\,i-j)\nonumber\\
&&-\sum_{l}t_{il}\langle\hat{T}n^f_{i}(\tau)d_l(\tau)d^{\dag}_{j}(0)\rangle=n^f\delta_{ij}\,\delta(\tau)\,,
\end{eqnarray}
where $n^f$ is the averaged number of the localized electrons. In the Hubbard-I approximation, the following decoupling in the second term in \eqref{Fd} is made~\cite{hubbard1}: $\langle\hat{T}n^f_{i}(\tau)d_l(\tau)d^{\dag}_{j}(0)\rangle\to-n^fG_d(\tau,\,i-j)$. Under this substitution, the closed system of equations for the $G_d$ and $F_d$ is obtained. As a result, one obtains the following expression for the one-particle Green's function (in the frequency-momentum representation):
\begin{equation}\label{GdHI}
G_d(i\omega_n,\mathbf{k})=\frac{i\omega_n+\mu-U(1-n^f)}{\left[i\omega_n+\mu-E_{+}(\mathbf{k})\right]\left[i\omega_n+\mu-E_{-}(\mathbf{k})\right]},
\end{equation}
where $\omega_n=(2n+1)\pi T$ are the Matsubara frequencies,
\begin{equation}\label{HIspec}
E_{\pm}(\mathbf{k})=\frac{\varepsilon(\mathbf{k})+U}{2}\pm\sqrt{\left(\frac{\varepsilon(\mathbf{k})-U}{2}\right)^2+Un^f\varepsilon(\mathbf{k})}\,,
\end{equation}
and $\varepsilon(\mathbf{k})=-\sum_jt_{ij}e^{\mathbf{k}(\mathbf{r}_i-\mathbf{r}_j)}$ is the spectrum of free $d$-electrons. The band structure of interacted $d$-electrons consists of two Hubbard subbands with the spectra $E_{\pm}(\mathbf{k})$. The filling of lower (upper) subband is equal to $1-n^f$ ($n^f$). The gap between Hubbard subbands exists for any value of $U$. This is the main weakness of the Hubbard-I approximation: it can not explain the metal to insulator transition existed in the Hubbard-like models. The one-particle Green's function of the $f$-electrons $G_f(\tau-\tau',\,i-j) = -\langle\hat{T}f_i(\tau)f^{\dag}_{j}(\tau')\rangle$ is obtained in the same manner. The result for $G_f$ in the Hubbard-I approximation is
\begin{equation}\label{Gf}
G_f(i\omega_n,\mathbf{k})=\frac{1-n^d}{i\omega_n+\mu+\epsilon_f}+\frac{n^d}{i\omega_n+\mu+\epsilon_f-U}\,.
\end{equation}
The $f$-electrons Green's function remains local, but the localized level splits into two sublevels corresponding to sites occupied by or free of $d$-electrons. The filling of the lower (upper) level is equal to $1-n^d$ ($n^d$).

The knowledge of one-particle Green's functions allows to evaluate the thermodynamics quantities of the system. The average number of $d$-electrons is given by the relation $n^d=T\sum_n\int d\mathbf{k}/V_{BZ}e^{i\omega_n0}G_d(i\omega_n,\mathbf{k})$, where $V_{BZ}$ is the volume of the first Brillouin zone. Taking the summation over Matsubara frequencies, after straightforward algebra, one obtains
\begin{equation}\label{nd}
n^d=\sum_{s=\pm}\int\limits_{E^{\rm{min}}_{s}}^{E^{\rm{max}}_{s}}\!\!\!\! d E\,\rho_0\!\!\left(\frac{E(E-U)}{E-U(1-n^f)}\right)n_{F}(E)\,,
\end{equation}
where $E^{\rm{max}}_{\pm}$ and $E^{\rm{min}}_{\pm}$ are the maximum and minimum values of the functions $E_{\pm}(\mathbf{k})$,
\begin{equation}
n_{F}(E)=\frac{1}{e^{\beta(E-\mu)}+1}\,,\;\;\beta=\frac{1}{T}\,,
\end{equation}
and $\rho_0(E)$ is the density of states of free $d$-electrons
\begin{equation}
\rho_0(E)=\int\frac{ d\mathbf{k}}{V_{BZ}}\delta(E-\varepsilon(\mathbf{k}))\,.
\end{equation}
This function is non-zero only in the region $|E|\leq W/2$, where $W$ is the bandwidth of free $d$-electrons. Taking into account that $\rho_0(E)=0$ outside this region, one can extend the integration over $E$ in \eqref{nd} to all real axis, and omit the summation over two Hubbard subbands.

The average number of $f$-electrons is
\begin{equation}\label{nf}
n^f=(1-n^d)n_F(-\epsilon_f)+n^dn_F(U-\epsilon_f)\,.
\end{equation}
When  the total number $n$ of electrons is fixed, the chemical potential is determined from the relation $n=n^d(\mu)+n^f(\mu)$. This equation, and equations~\eqref{nd} and~\eqref{nf} set the system of equation for the self-consistent determination of $\mu$, $n^d$, and $n^f$. For the purposes of the present study, it is more convenient to consider the system at fixed value of $n^f$. In this case, the chemical potential $\mu$ is found from \eqref{nf}. At small temperatures, $\mu$ becomes fixed at the point $\mu_0=-\epsilon_f$ when $n<1$, or at $\mu_0=-\epsilon_f+U$ when $n>1$.

The density of states for (interacted) $d$-electrons in the Hubbard-I approximation is
\begin{equation}\label{DOSHI}
\rho(E)=\rho_0\!\!\left(\frac{E(E-U)}{E-(1-n^f)U}\right)\,.
\end{equation}
The free energy of the system $F$ can be written in the form
\begin{eqnarray}\label{FHI}
F&=&\int\limits_{-\infty}^{\infty}\!\! d E\,\rho(E)\,\Phi(E)+\nonumber\\
&&(1-n^d)\Phi(-\epsilon_f)+n^d\Phi(U-\epsilon_f)\,,
\end{eqnarray}
where
\begin{equation}
\Phi(E)=\mu\,n_F(E)-T\ln\left[1+e^{\beta(\mu-E)}\right]\,.
\end{equation}
The one-particle Green's functions and the mean values $n^d$ and $n^f$ as functions of $n$ will be compared in the next sections with those calculated using DMFT approximation.

\section{Falicov-Kimball model in DMFT approximation}\label{SecDMFT}

In the dynamical mean field approach, the lattice problem is replaced by the effective single-site problem. The interaction of electrons in a given site, say $i=o$, with other sites in the lattice is described by the effective dynamical field $\lambda(\tau)$ which has to be determined self-consistently. For the Falicov-Kimball model, the single-site Hamiltonian is
\begin{equation}\label{H0}
H_0=-\epsilon_{f}n^{f}+Un^{d}n^{f}-\mu(n^{d}+n^{f}),
\end{equation}
where $n^d=d^{\dag}d$, $n^f=f^{\dag}f$, and $d^{\dag}$, $f^{\dag}$ ($d$, $f$) are creation (annihilation) operators for $d$- and $f$-electrons at site $i=o$. The partition function ${\cal Z}$ for this single-site model can be represented as a functional integral of $\exp(-S_{\rm{eff}})$ with the effective action $S_{\rm{eff}}=\int_0^{\beta} d\tau(d^{\dag}\partial d/\partial\tau+f^{\dag}\partial f /\partial\tau+H_0)$ + $\int_0^{\beta} d\tau \int_0^{\beta} d\tau'$ $d^{\dag}(\tau)\lambda(\tau-\tau')d(\tau')$. The partition function can be also written in the interaction representation with the non-local in time interaction operator $U_{\rm{int}}(\tau)$ = $\int_0^{\beta} d\tau'd^{\dag}(\tau)\lambda(\tau-\tau')d(\tau')$. Both representations are, of course, equivalent. In general, for the Hubbard-like models, the partition function can be calculated only numerically since $H_0$ is not quadratic in electronic operators. Fortunately, for the Falicov-Kimball model, it is possible to obtain an analytical formula for ${\cal Z}$. This is related to the fact that $n^f$ commutes both with $H_0$ and $U_{\rm{int}}$.

In the interaction representation the partition function has a form~\cite{Brandt}
\begin{equation}\label{Z}
{\cal Z}=\Tr\left[e^{-\beta H_0}\sigma\right],
\end{equation}
where
\begin{equation}\label{sigma}
\sigma=\hat{T}\exp\left[-\int_0^{\beta}\!\!\!\! d\tau\!\!\int_0^{\beta}\!\!\!\! d\tau'd^{\dag}(\tau)\lambda(\tau-\tau')d(\tau')\right],
\end{equation}
and
\begin{equation}
d(\tau)=e^{\tau H_0}de^{-\tau H_0}\,,\;\;d^{\dag}(\tau)=e^{\tau H_0}d^{\dag}e^{-\tau H_0}\,.
\end{equation}
Since the operator $n^f$ commutes both with $d(\tau)$ and $H_0$, the trace over $f$ operators is trivial and gives
\begin{equation}\label{Z0}
{\cal Z}=\Tr\left[e^{\beta\mu n^d}\sigma\right]+e^{\beta(\mu+\epsilon_f)}\Tr\left[e^{\beta(\mu-U)n^d}\sigma\right],
\end{equation}
where traces are taken over $d$ operators only. Expanding each term in perturbation expansion series, taking the traces by applying the Wick theorem, and re-summing the series again, one obtains:
\begin{equation}\label{Zn}
{\cal Z}=\left[1+e^{\beta\mu}\right]\!\!Z_{\lambda}(\mu)+e^{\beta(\mu+\epsilon_f)}\!\!\left[1+e^{\beta(\mu-U)}\right]\!\!Z_{\lambda}(\mu-U),
\end{equation}
where
\begin{equation}\label{Zlambda}
Z_{\lambda}(E)=\prod_{n=-\infty}^{\infty}\!\!\!\left(1-\frac{\lambda_n}{i\omega_n+E}\right),
\end{equation}
and $\lambda_n=\int_0^{\beta} d\tau e^{-i\omega_n\tau}\lambda(\tau)$ is the Fourier transform of the function $\lambda(\tau)$.

The local $d$-electrons Green's function is
\begin{eqnarray}\label{Gdlocal}
&&D(i\omega_n)=-\frac{1}{{\cal Z}}\frac{\delta{\cal Z}}{\delta\lambda_n}=\frac{1-n^f}{i\omega_n+\mu-\lambda_n}+\\
&&\frac{n^f}{i\omega_n+\mu-\lambda_n-U}\equiv\frac{1}{i\omega_n+\mu-\lambda_n-\Sigma_n}\,,\nonumber
\end{eqnarray}
where $\Sigma_n=\Sigma(i\omega_n)$ is the self energy of the effective single-site model, and $n^f$ is the averaged number of $f$-electrons
\begin{eqnarray}\label{nfDMFT}
n^f&=&\frac{T}{{\cal Z}}\frac{\delta{\cal Z}}{\delta\epsilon_f}=\\
&&\left[1+e^{-\beta(\mu+\epsilon_f)}\frac{Z_{\lambda}(\mu)}{Z_{\lambda}(\mu-U)}\frac{1+e^{\beta\mu}}{1+e^{\beta(\mu-U)}}\right]^{-1}.\nonumber
\end{eqnarray}
The average number of $d$-electrons is
\begin{equation}\label{ndDMFT}
n^d=\frac{1}{\beta}\sum_{n}D(i\omega_n)e^{i\omega_n0}=\frac12+\frac{1}{\beta}\sum_{n}D(i\omega_n)\,.
\end{equation}

The $d$-electrons Green's function in the lattice can be written in the form
\begin{equation}\label{GdSigma}
G_d(i\omega_n,\mathbf{k})=\frac{1}{i\omega_n+\mu-\varepsilon(\mathbf{k})-\Sigma(i\omega_n,\mathbf{k})}\,,
\end{equation}
where $\Sigma(i\omega_n,\mathbf{k})$ is the self energy of the lattice model. In the DMFT approximation, it is assumed that the self energy $\Sigma(i\omega_n,\mathbf{k})$ does not depend on the quasimomentum and equals to $\Sigma(i\omega_n,\mathbf{k})=\Sigma_n$, with $\Sigma_n$ from \eqref{Gdlocal}. This is uncontrollable scheme which becomes exact, nevertheless, in infinite dimensions~\cite{DMFTreview}. From \eqref{GdSigma}, one obtains
\begin{equation}\label{Gdlocal1}
D(i\omega_n)=\int\!\!\frac{ d\mathbf{k}}{V_{BZ}}G_d(i\omega_n,\mathbf{k})\equiv\tilde{D}\left(i\omega_n+\mu-\Sigma_n\right)\,,
\end{equation}
where
\begin{equation}\label{DE}
\tilde{D}(E)=\int\limits_{-W/2}^{W/2}\!\!\!\! d E'\,\frac{\rho_0(E')}{E-E'}\,.
\end{equation}

Equations~\eqref{Gdlocal} and~\eqref{Gdlocal1} allow to calculate self-consistently both $\lambda_n$ and $\Sigma_n$. From \eqref{Gdlocal} and~\eqref{Gdlocal1} one obtains
\begin{eqnarray}
\lambda_n&=&i\omega_n+\mu-\Sigma_n-\frac{1}{\tilde{D}\left(i\omega_n+\mu-\Sigma_n\right)}\,,\label{sys1}\\
\Sigma_n&=&n^fU\frac{i\omega_n+\mu-\lambda_n}{i\omega_n+\mu-\lambda_n-(1-n^f)U}\,,\label{sys2}
\end{eqnarray}
where $n^f$ is given by \eqref{nfDMFT}. This system of equations is solved by iterations. In the iteration scheme it is more convenient to work  either at fixed number of localized electrons $n^f$ or at fixed total number of electrons $n$,  instead of fixed chemical potential $\mu$ (especially at low temperatures). For given  $\lambda_n$, $\Sigma_n$, and $\mu$, the new values of $\lambda_n$ are calculated from \eqref{sys1}, the new value of $\mu$ is found from \eqref{nfDMFT} if $n^f$ is fixed or from the relation $n=n^d(\mu)+n^f(\mu)$ if $n$ is fixed, and, finally, the new values of $\Sigma_n$ are evaluated from \eqref{sys2}. For the relative accuracy $10^{-4}$, this iteration procedure converges in about $10$ iterations.

\section{Comparison of Hubbard-I and DMFT approximations}\label{Comparison}

\begin{figure}[t]
\begin{center}
\includegraphics*[width=0.5\columnwidth]{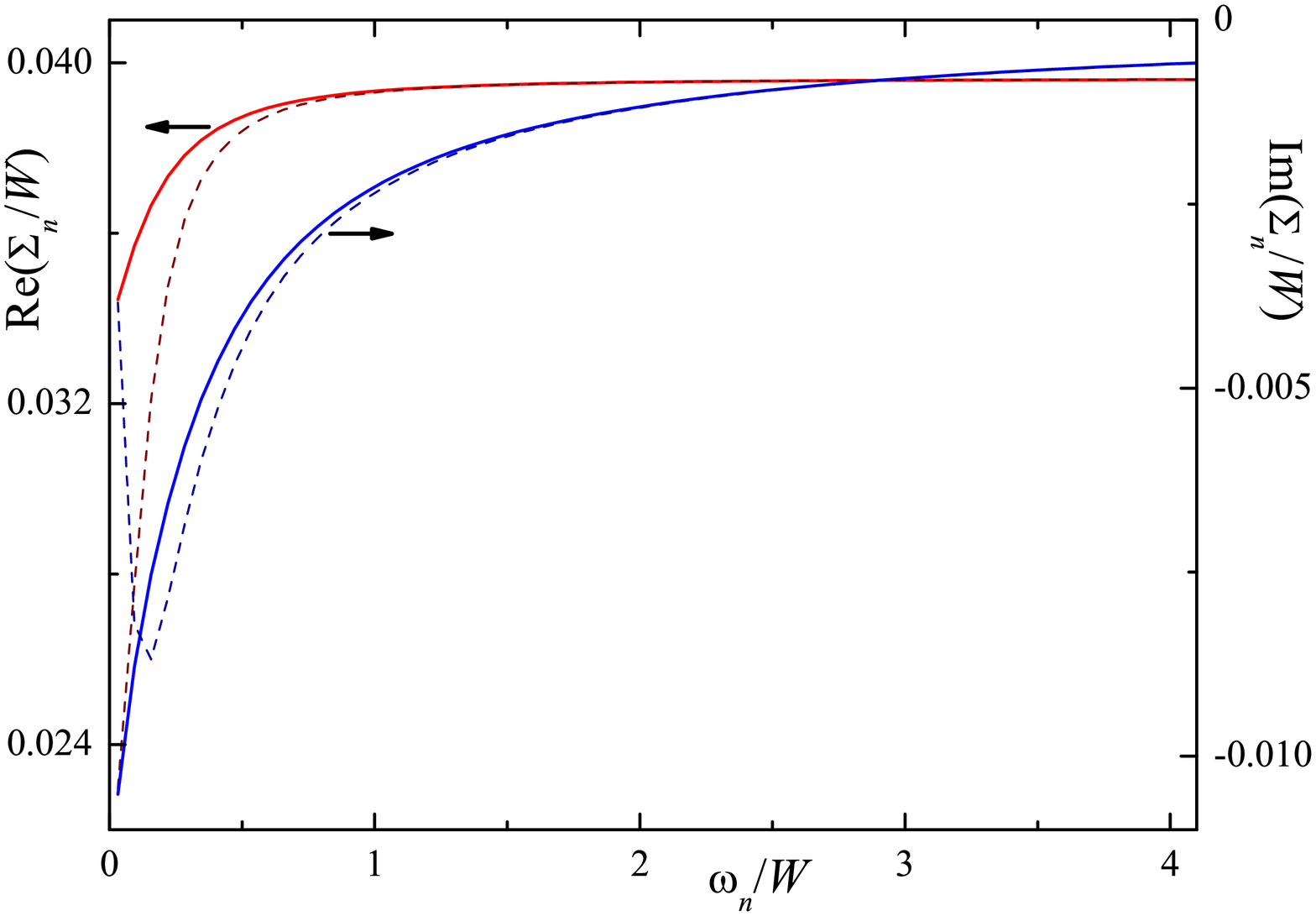}\\
\includegraphics*[width=0.5\columnwidth]{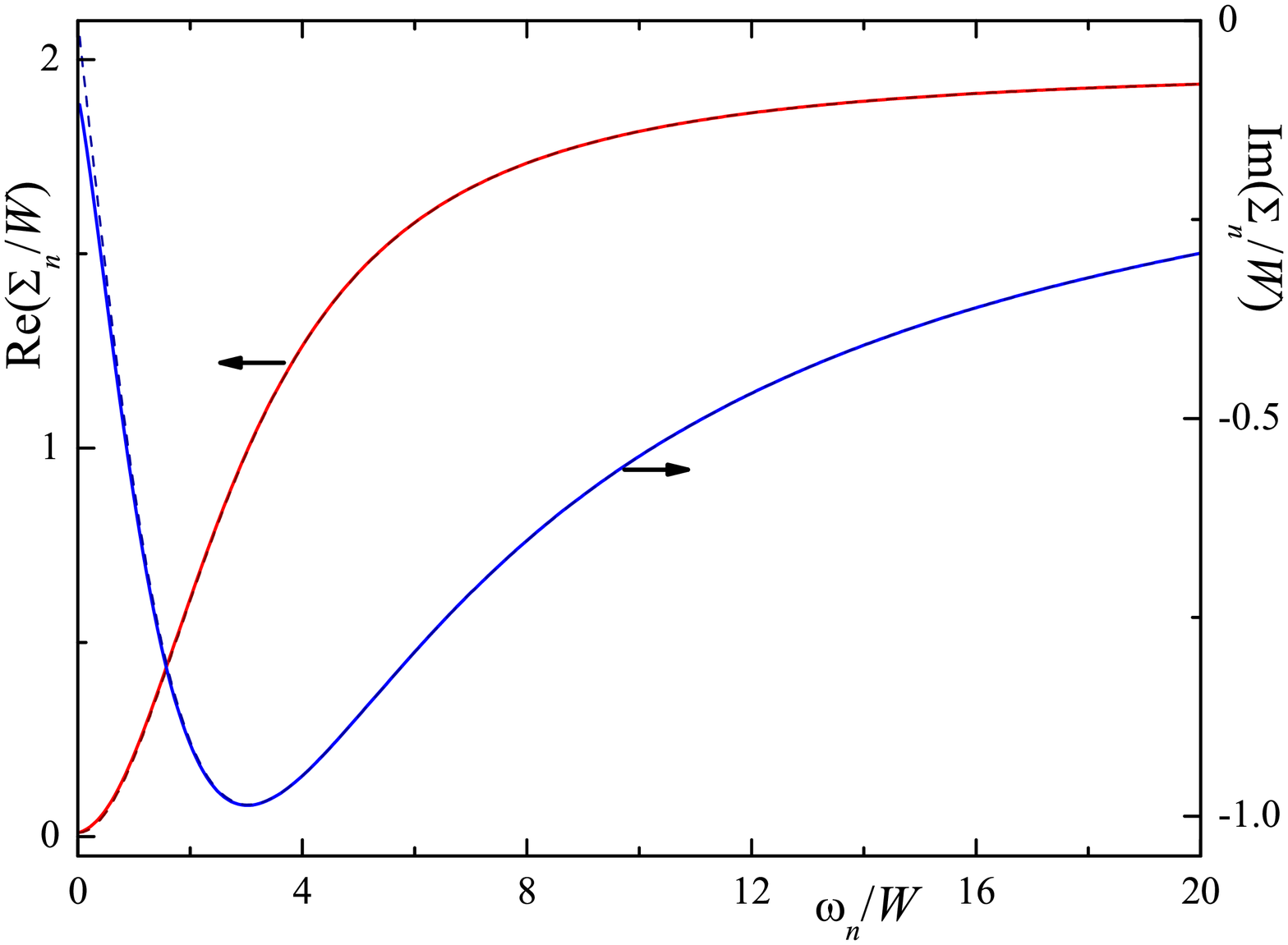}
\end{center}
\caption{\label{FigSigma}The dependencies of $\rm{Re}\,\Sigma_n$ and $\rm{Im}\,\Sigma_n$ vs $\omega_n$ calculated for $U/W=0.1$ (upper panel) and $U/W=5$ (lower panel). Other parameters are: $n^f=0.4$, $\epsilon_f/W=0.1$, $T/W=0.01$. Solid (dashed) curves correspond to the DMFT (Hubbard-I) approximation.}
\end{figure}

Consider first the itinerant electrons Green's function. All results below are obtained for the simple cubic lattice with the spectrum of free $d$-electrons of the form $\varepsilon(\mathbf{k})=-W(\cos k^x+\cos k^y+\cos k^z)/6$. In the DMFT approximation the $d$-electrons Green's function is given by \eqref{GdSigma}, where $\Sigma(i\omega_n,\mathbf{k})=\Sigma_n$ is calculated from the system of equations~\eqref{nfDMFT}, \eqref{sys1}, and \eqref{sys2}. The typical curves of real and imaginary parts of $\Sigma_n$ vs $\omega_n$ are shown in figure~\ref{FigSigma} for small ($U=0.1W$) and large ($U=5W$) values of $U$. From  \eqref{sys1} and~\eqref{sys2} it is possible to obtain an analytical expression for $\Sigma_n$ in the narrow band limit, when $W\to0$. Expanding the function $\tilde{D}(E)$ in the Taylor series up to the second order, from the system~\eqref{sys1} and \eqref{sys2} one obtains:
\begin{eqnarray}
\lambda_n\!\!&\approx&\!\!\frac{W^2}{24}\!\!\left[\frac{1-n^f}{i\omega_n+\mu}+\frac{n^f}{i\omega_n+\mu-U}\right]\!\!+\!\Or\!\!\left(\frac{W^4}{E^3}\right)\!,\label{sys01}\\
\Sigma_n\!\!&\approx&\!\!\Sigma^{(0)}_n+\Or\!\!\left(\frac{W^2}{E}\right),\nonumber\\
\Sigma^{(0)}_n\!\!&=&\!\!\frac{n^fU(i\omega_n+\mu)}{i\omega_n+\mu-(1-n^f)U}\,.\label{sys02}
\end{eqnarray}
Substituting \eqref{sys02} into \eqref{GdSigma}, after simple algebra, one arrives to \eqref{GdHI} for the Green's function in the Hubbard-I approximation. The self energy $\Sigma^{(0)}_n$ corresponding to the Hubbard-I approximation is also shown in figure~\ref{FigSigma}. It is seen that at small $U$ the asymptotic~\eqref{sys02} has qualitatively different behavior in the region $\omega_n\lesssim W$ than $\Sigma_n$ calculated from \eqref{sys1} and \eqref{sys2}. For large $U$ both curves practically coincide even for small values of $\omega_n$.

Thus, both approximations give similar results for large $U$. Note, however, that limits $U\gg W$ and $W\to0$ are different, and even at $U\to\infty$ there will be a discrepancy between $\Sigma_n$  and $\Sigma^{(0)}_n$ at small frequencies $\omega_n\ll W$. One can say, that the Hubbard-I approximation corresponds to the DMFT one in the narrow band limit $W\to0$.

The use of Matsubara Green's functions is convenient for calculation of the thermodynamics quantities. But if we are interesting in the spectral properties, we need working at real frequencies. There are two possible ways for transition from Matsubara to real frequencies. One of them is the analytical continuation of calculated Matsubara Green's function to real axis. It can be done by different methods~\cite{DMFTreview,maxent,pade}.
In this paper I use Pad\'{e} approximants method~\cite{pade} in evaluating the real-frequency Green's function.
Another way I propose in this paper is the use of analytical continuation of the system of equations~\eqref{sys1} and~\eqref{sys2} which is obtained by replacing $i\omega_n\to\omega+i0$. The infinitesimal positive imaginary part is added to calculate the retarded self-energy. The system of equations~\eqref{sys1} and~\eqref{sys2} becomes
\begin{eqnarray}
\lambda(\omega)&=&\omega+\mu-\Sigma(\omega)-\frac{1}{D_{\rm R}\left(\omega+\mu-\Sigma(\omega)\right)}\,,\label{sys1real}\\
\Sigma(\omega)&=&n^fU\frac{\omega+\mu-\lambda_n}{\omega+\mu-\lambda(\omega)-(1-n^f)U}\,,\label{sys2real}
\end{eqnarray}
where $D_{\rm R}(E)=\tilde{D}(E+i0)$, and for real $E$, $D_{\rm R}(E)=\tilde{D}(E)-i\pi\rho_0(E)$. The values of $\mu$ or/and $n^f$ in \eqref{sys1real} and~\eqref{sys2real} are found by solving the iteration scheme at Matsubara frequencies as described above. Excluding the $\lambda(\omega)$ from the system~\eqref{sys1real} and~\eqref{sys2real}, one arrives to the equation for the self-energy:
\begin{equation}\label{SigmaReal}
\frac{\Sigma(\omega)\left[\Sigma(\omega)-U\right]}{\Sigma(\omega)-n^fU}=-\frac{1}{D_{\rm R}\left(\omega+\mu-\Sigma(\omega)\right)}\,.
\end{equation}
This equation is solved numerically. Such method of finding the retarded self-energy is more reliable than the Pad\'{e} approximants method of analytical continuation of $\Sigma_n$, especially at high frequencies, where Pad\'{e} approximants method depends on the cut-off frequency and can fail in some cases. But at frequencies $|\omega|\lesssim W$ both methods give the same results.

\begin{figure}[t]
\begin{center}
\includegraphics*[width=0.35\columnwidth]{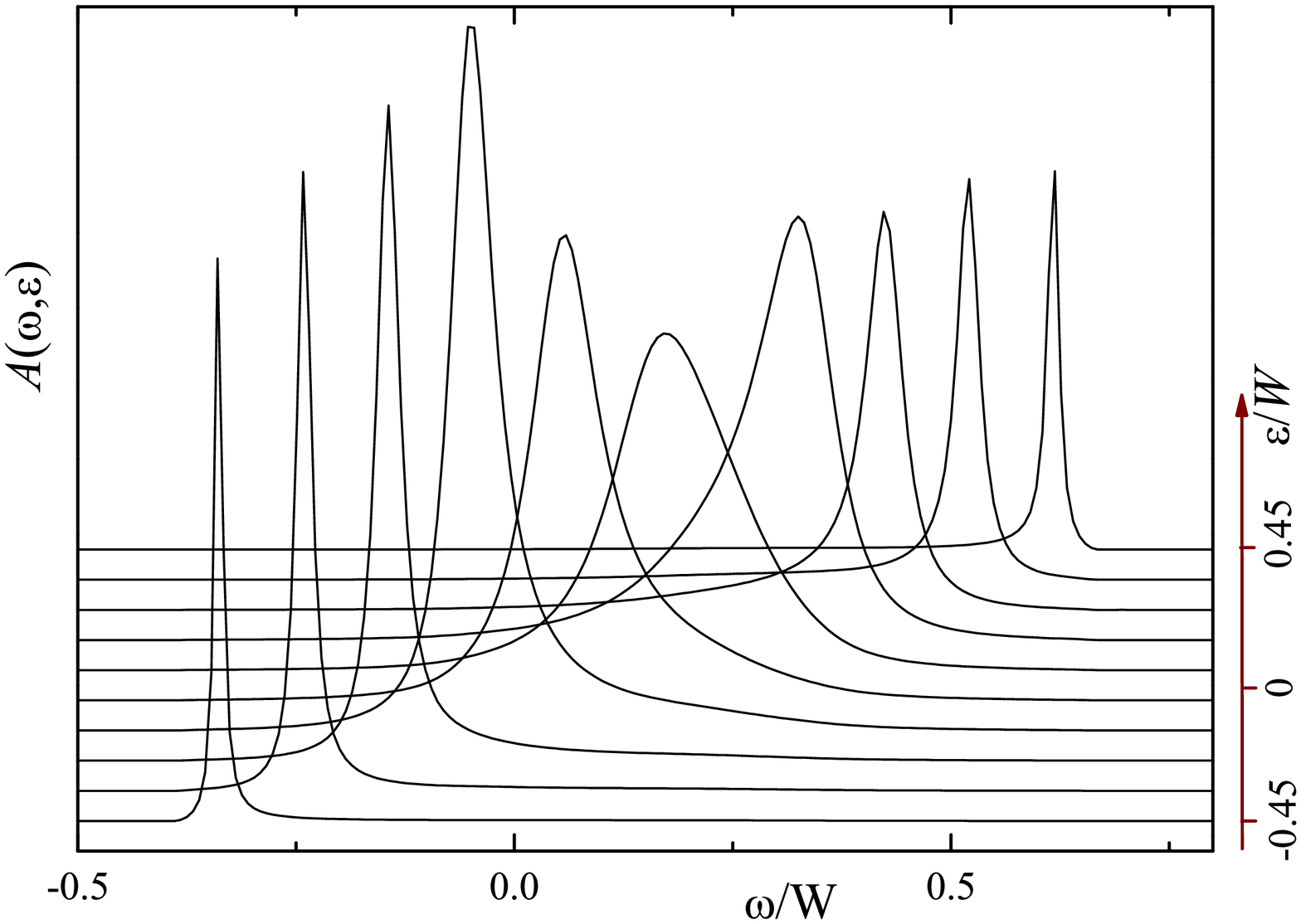}\\
\includegraphics*[width=0.35\columnwidth]{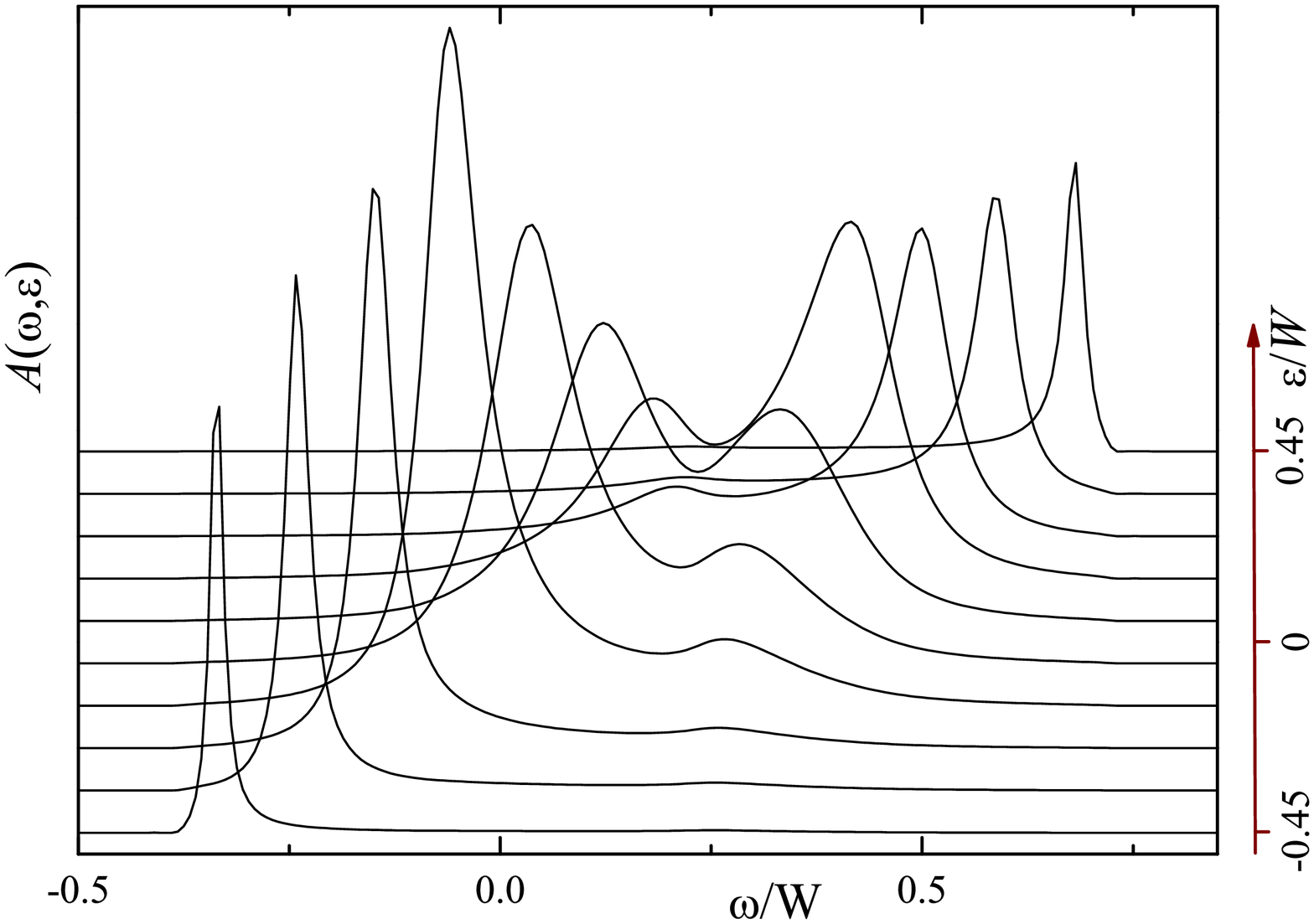}\\
\includegraphics*[width=0.35\columnwidth]{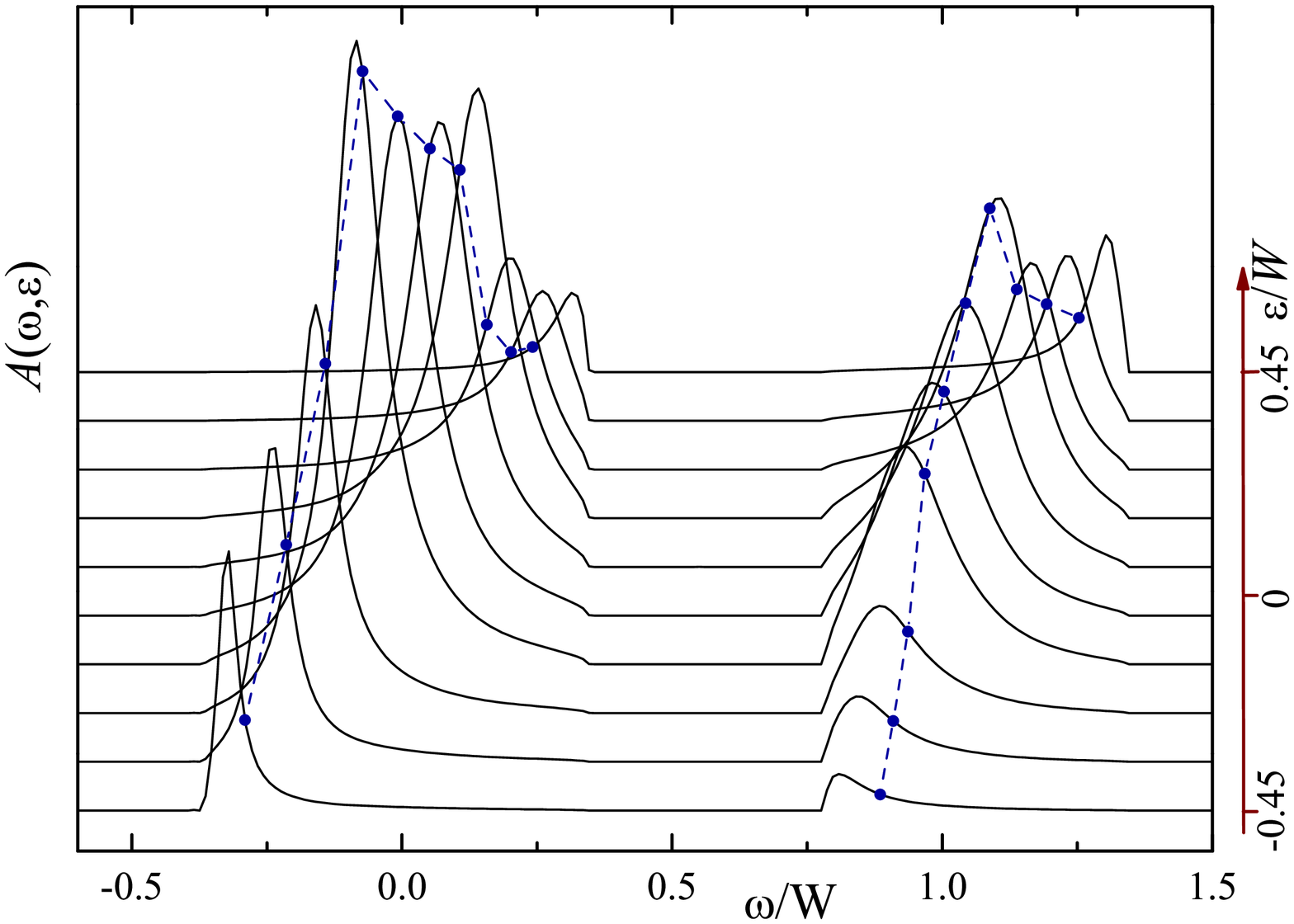}\\
\end{center}
\caption{\label{FigA}The frequency dependence of the spectral function \eqref{A} at different values of $\varepsilon$ and for three values of $U$: $U/W=0.1$, $U/W=0.3$, and $U/W=1$ for panels from top to bottom. Other model parameters are: $n^f=0.4$, $\epsilon_f/W=0.1$, and $T/W=0.01$. In the lowest panel, circles connected by dashed lines mark the position of the $\delta$-peaks of the spectral function calculated in the Hubbard-I approximation.}
\end{figure}

It is possible to obtain an analytical expression for $\Sigma(\omega)$ in two different limits: for small $U$, and for small number of $f$-electrons, $n^f\ll1$. In the former case, expanding $\lambda$ and $\Sigma$ in \eqref{sys1real} and~\eqref{sys2real} in series of $U$, one obtains in the second order:
\begin{eqnarray}
\Sigma(\omega)&=&n^fU+n^f(1-n^f)U^2D_{\rm R}(\omega+\mu)\,,\\
{\rm Im}\,\Sigma(\omega)&=&-\pi n^f(1-n^f)U^2\rho_0(\omega+\mu)\,.\nonumber
\end{eqnarray}
This result coincides with $\Sigma$ calculated by perturbation theory in the second order of $U$. Note a non-Fermi liquid behavior of the self-energy~\cite{NonFL,DMFTreview}: $\rm{Im}\,\Sigma(0)\neq0$ even for $T\to0$. This is due to the infinite mass of $f$-electrons. If we allow of $f$-electrons to have a finite bandwidth $W_f$, then the perturbation theory gives the standard Fermi-liquid result for the imaginary part of the self-energy at frequencies $|\omega|\ll W_f$:
\begin{equation}
{\rm Im}\,\Sigma(\omega)\propto-(\omega^2+T^2)\,.\nonumber
\end{equation}

\begin{figure}
\begin{center}
\includegraphics*[width=0.5\columnwidth]{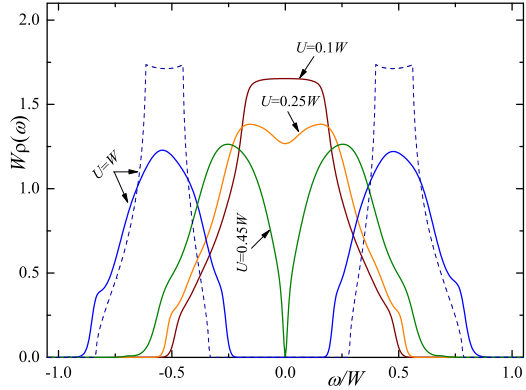}\\
\end{center}
\caption{\label{FigDOS}The density of states in DMFT (solid curves) and Hubbard-I (dashed curve) approximation calculated at different values of $U$. Other model parameters are: $n^f=0.5$, $\epsilon_f=0$, $T/W=0.01$.}
\end{figure}

At $n^f\ll1$, the expansion of $\lambda$ and $\Sigma$ in the series of $n^f$ gives in the leading order:
\begin{equation}\label{SigmaNf}
\Sigma(\omega)=\frac{n^fU}{1-U\,D_{\rm R}(\omega+\mu)}\,.
\end{equation}
This is a well known result for the self-energy of the electrons scattered on localized impurities in the limit of small concentration of impurities~\cite{Mahan}. Thus, in both limits, the DMFT approximation gives results confirming by controllable approximations such as perturbation theory.


Consider now the spectral properties of the conduction electrons. The spectral function of $d$-electrons is $A(\omega,\mathbf{k})=-{\rm Im}\,G_d(\omega,\mathbf{k})/\pi$. Since in both approximations all momentum dependence contains in the non-interacted spectrum $\varepsilon(\mathbf{k})$, it is more convenient to consider the function
\begin{equation}\label{A}
A(\omega,\varepsilon)=-\frac{1}{\pi}\,{\rm Im}\frac{\rho_0(\varepsilon)}{\omega+\mu-\varepsilon-\Sigma(\omega)}\,.
\end{equation}
The curves of $A(\omega,\varepsilon)$ vs $\omega$ at different values of $\varepsilon$ are shown in figure~\ref{FigA} for three different values of the interaction constant $U$. At small $U$, there is only one peak in the spectral function for any value of $\varepsilon$. In an intermediate case, the two-peaked structure of the spectral function appears at some range of $\varepsilon$. Finally, at large $U$, there exists two separate Hubbard bands of $d$-electrons. The self-energy in the Hubbard-I approximation has no imaginary part, the spectral function consists of two $\delta$-peaks located at frequencies $\omega=E_{\pm}(\varepsilon)-\mu$, where $E_{\pm}$ are given by \eqref{HIspec} with $\varepsilon(\mathbf{k})=\varepsilon$. The position of these peaks are shown by circles in the lowest panel of figure~\ref{FigA} corresponding to $U=W$. It is seen from this figure, that both approximations give very similar results for the band structure of $d$-electrons in the large $U$ limit (with the exception of finite lifetime of quasiparticles in DMFT).

The $d$-electrons density of states $\rho(E)$ is obtained by integration of $A(E-\mu,\varepsilon)$ over $\varepsilon$. Using \eqref{SigmaReal}, from \eqref{A} one obtains:
\begin{equation}\label{DOSdmft}
\rho(E)=\frac{1}{\pi}\,{\rm Im}\left(\frac{1-n^f}{\Sigma(E-\mu)-U}+\frac{n^f}{\Sigma(E-\mu)}\right)\,.
\end{equation}
The density of states \eqref{DOSdmft} calculated for the symmetric case, $\epsilon_f=0$, $n^f=0.5$, and at different values of $U$ are shown in figure~\ref{FigDOS}. The density of states \eqref{DOSHI} in the Hubbard-I approximation is shown for $U=W$. In this case, the lower Hubbard subband turns out to be filled, $n^d=0.5$. It is seen, that in the DMFT approach the metal-insulator transition occurs at $U_c\cong0.45W$ (in the symmetric case, $\epsilon_f=0$, and for simple cubic lattice). Below, I will study the case $U>U_c$, where both approximations give qualitatively the same results for the density of states.

\section{Phase Separation}\label{PS}

In the previous sections, the homogeneous state of the Falicov-Kimball model was studied. But it is well known that for wide range of the model parameters an inhomogeneous state is preferable at low temperatures~\cite{FKreview}. It can be either charge density wave, or phase separation. For the symmetric case ($\varepsilon_f=0$) and at half-filling, the checkerboard phase is realized~\cite{Brandt,GruberCDW,FreericksCDW_PS}, while away of half-filling the phase separation prevails~\cite{FreericksCDW_PS,Maska,FreericksPS,WePRL}.

\begin{figure}[t]
\begin{center}
\includegraphics*[width=0.5\columnwidth]{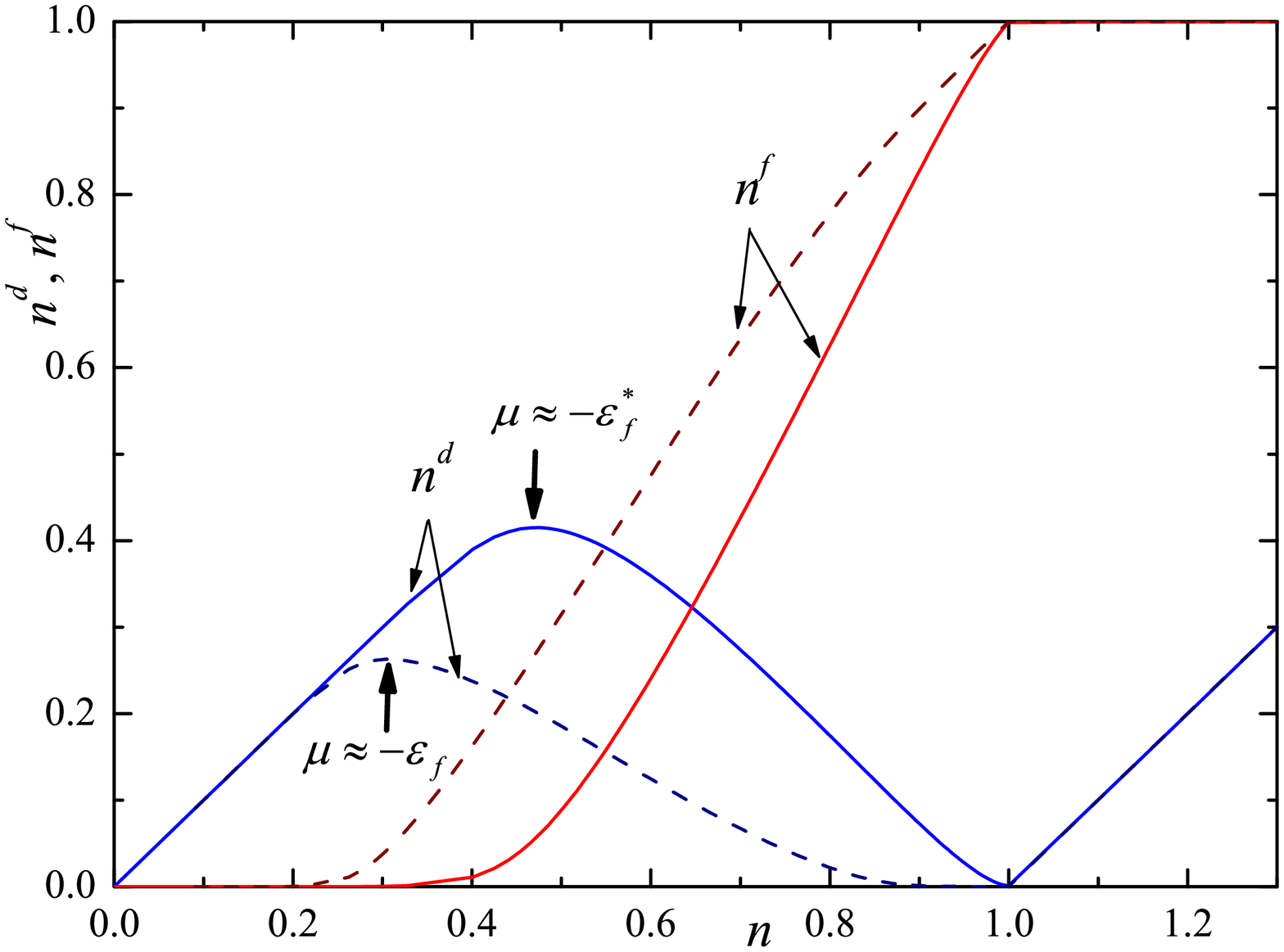}\\
\end{center}
\caption{\label{FigN}The dependencies of $n^d$ and $n^f$ on $n$. Solid (dashed) curves correspond to the DMFT (Hubbard-I) approximation. The parameters of the model are: $U/W=1.5$, $\epsilon_f/W=0.1$, $T/W=0.01$. The vertical arrows show the position of doping $n$ when the chemical potential equals to the bare and renormalized energy of the localized level (see the text below).}
\end{figure}

In this paper I study non-symmetric case ($\varepsilon_f\neq0$) and restrict myself by consideration of the phase separation only. To do this, I consider the evolution of the system with doping and examine the stability of the homogeneous state by analyzing the system's free energy. The typical dependencies of numbers of $d$- and $f$-electrons, $n^d$ and $n^f$, on doping $n$ are shown in figure~\ref{FigN}. Both approximations give qualitatively the same behavior of $n^d$ and $n^f$ vs $n$. At small doping, $n\ll1$, only $d$-electrons exist in the system since the chemical potential lies far below the localized level. With the increase of $n$ the $f$-electrons appear. But the filling of correlated band depends on the number $n^f$ of localized electrons: both the bandwith and the filling of the lower Hubbard band decrease when $n^f$ increases. Thus, in the further increase of $n$ (and $n^f$), the number of itinerant electrons starts to decrease. Finally, at $n\gtrsim1$, all localized levels become occupied, and $d$-electrons in the upper Hubbard band appear.

\begin{figure}
\begin{center}
\includegraphics*[width=0.5\columnwidth]{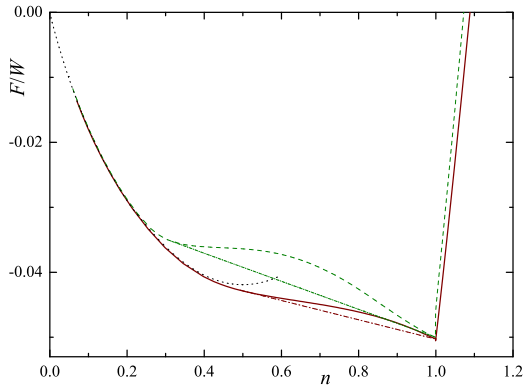}\\
\end{center}
\caption{\label{FigF}The free energy of the system vs doping. Solid (dashed) curve corresponds to the DMFT (Hubbard-I) approximation. Dotted curve is the free energy of free $d$-electrons. In the region where $d$- and $f$-electrons coexist the free energy has negative curvature indicating the instability of the homogeneous state. Maxwell construction to both curves is shown by dotted-dashed lines. The parameters of the model correspond to figure~\ref{FigN}.}
\end{figure}

The quantitative disagreement between dependencies  $n^d(n)$, $n^f(n)$ is related mainly to the different behavior of $n^f$ on the chemical potential $\mu$ in two approximations. For the Hubbard-I the dependence of $n^f$ on $\mu$ is governed by \eqref{nf}. In the case $U\gg W$ and $\mu<U$, it can be approximated as
\begin{equation}\label{nf0}
n^f\approx\frac{1-n^d}{e^{-\beta(\varepsilon_f+\mu)}+1}\,.
\end{equation}
Thus, at small temperatures, $f$-electrons appear when chemical potential reaches the value $\mu=-\varepsilon_f$. In the DMFT approximation, the dependence of $n^f$ on $\mu$ is given by \eqref{nfDMFT}. When $n^f\ll1$, one obtains from the system~\eqref{sys1} and~\eqref{sys2}: $\lambda_n\approx i\omega_n+\mu-1/\tilde{D}(i\omega_n+\mu)$. At $T\to0$, one can replace the Matsubara frequency summation by integration over $\omega$, and as a result, one obtains:
\begin{equation}\label{nf0DMFT}
n^f\approx\frac{1}{e^{-\beta[\varepsilon^*_f(\mu)+\mu]}+1}\,,
\end{equation}
where
\begin{equation}\label{DeltaEpsilon}
\varepsilon^*_f(\mu)=\varepsilon_f-f_0(\mu,\mu)+f_0(\mu-U,\mu)\,,
\end{equation}
and
\begin{equation}\label{F0}
f_0(E,\mu)=E\Theta(E)+\int\limits_0^{\infty}\frac{ d\omega}{\pi}\ln\left|\frac{E-\mu+1/\tilde{D}(i\omega+\mu)}{i\omega+E}\right|\,,
\end{equation}
Thus, in the DMFT approximation, the interaction between electrons renormalizes the energy of the localized level~\cite{NonFL}. The $f$-electrons appear when the chemical potential reaches the value defined by the equation $\mu+\varepsilon_f^*(\mu)=0$. Since the renormalized level has larger energy, the DMFT approximation gives the larger value of doping $n$ when $f$-electrons appear (and, consequently, the doping when $n^d$ starts to decrease, see figure~\ref{FigN}). With this exception, both approximations give the same behavior of $n^d$ and $n^f$ vs $n$.

\begin{figure}[t]
\begin{center}
\includegraphics*[width=0.5\columnwidth]{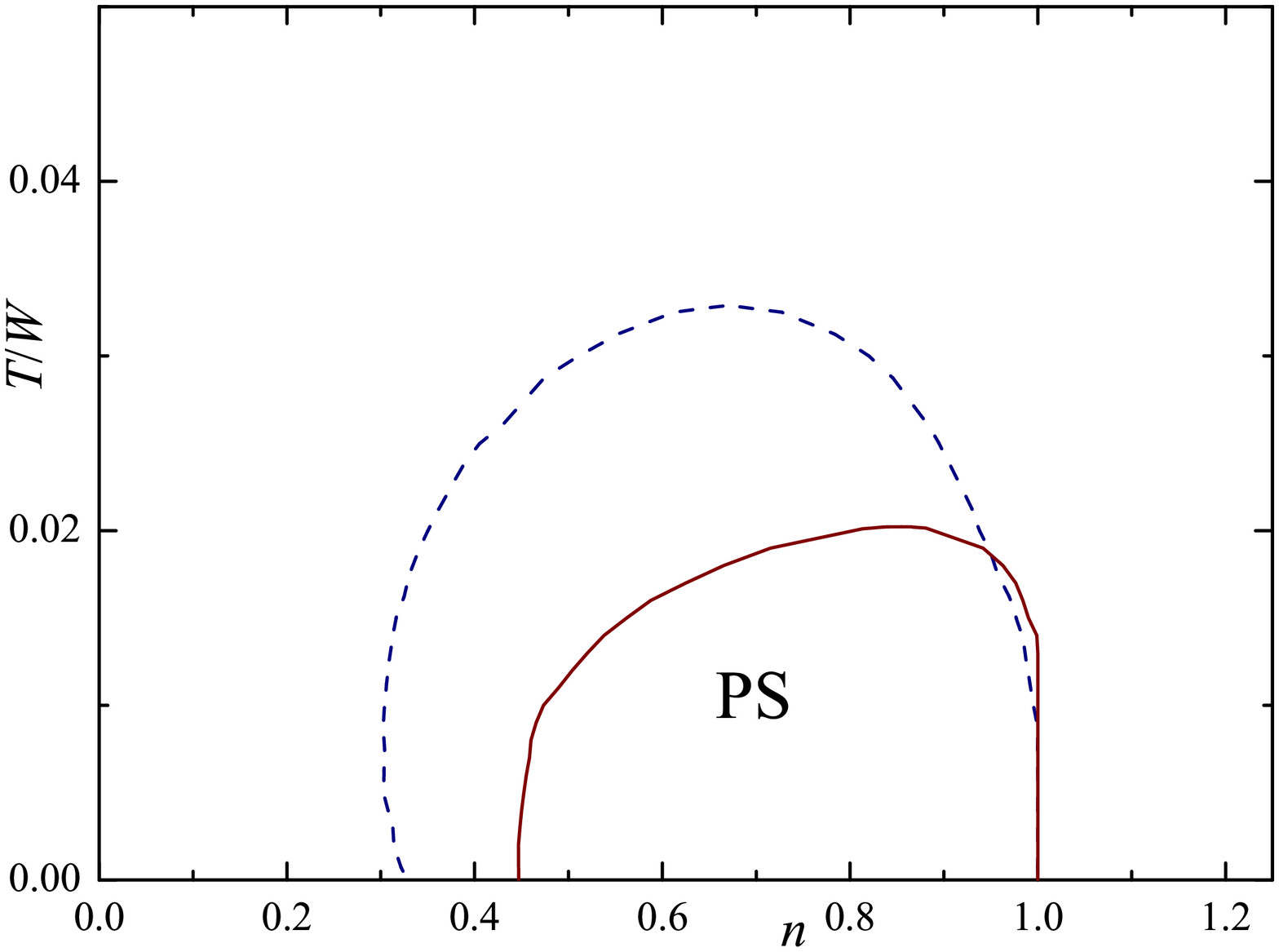}\\
\end{center}
\caption{\label{FigPhaseDiag} The phase diagram of the Falicov-Kimball model at $U/W=3$ and $\varepsilon_f/W=0.1$. Inside the closed region the phase separation exists. Solid (dashed) curve is calculated in the DMFT (Hubbard-I) approximation.}
\end{figure}

The free energy in the Hubbard-I approximation is given by \eqref{FHI}. In the DMFT approximation, the partition function \eqref{Zn} defines the grand potential of the effective single-site model: $\Omega_{loc}=-T\ln{\cal Z}$. The grand potential $\Omega$ and the free energy $F=\Omega+\mu n$ of the lattice model can be found following the approach described in \cite{Brandt}. As a result, one obtains:
\begin{equation}\label{FDMFT}
F=\Omega_{loc}+\mu n+T\sum_n\!\!\int\limits_{-\infty}^{\infty}\!\! d\varepsilon\,\rho_0(\varepsilon)\ln\!\left[\frac{i\omega_n+\mu-\lambda_n-\Sigma_n}%
{i\omega_n+\mu-\varepsilon-\Sigma_n}\right].
\end{equation}
The dependence of the system's free energy on doping calculated in two approximations is shown in figure~\ref{FigF}. The model parameters correspond to figure~\ref{FigN}. The free energies in two approximations coincide in the regions of $n$ where $n^f\ll1$ and at $n\approx1$ (where $n^d\ll1$). In the region of doping where itinerant and localized electrons coexist, the free energy has negative curvature indicating the instability of the homogeneous state. In this region the system separates into the phases with $n^f\approx0$ and $n^f\approx1$. With the increase of temperature, the doping range $n_1<n<n_2$ where phase separation exists becomes narrow and disappears at some critical temperature $T_c$. This result is summarized in the phase diagram presented in figure~\ref{FigPhaseDiag}. The Hubbard-I approximation gives the value of $n_1$ lower than the DMFT one. This is related to the fact that in the DMFT approximation $f$-electrons appear at larger doping due to renormalization of the energy of the localized level as described above. In addition, the DMFT gives the lower critical temperature.

\section{Extension to the Hubbard model}

Commutativity of the Falicov-Kimball model's Hamiltonian with the number of $f$-electrons strongly simplifies the DMFT scheme for this model; it is possible to obtain an analytical expression for the single-site partition function ${\cal Z}$. For other models of strongly correlated electrons, such as Hubbard model, the DMFT approach requires much more CPU time. The partition function ${\cal Z}$ and the local Green's functions can be calculated either numerically or by using some approximations which work only in the special ranges of model parameters.

\begin{figure}
\begin{center}
\includegraphics*[width=0.5\columnwidth]{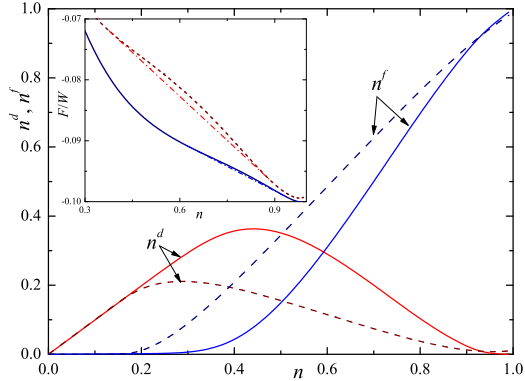}\\
\end{center}
\caption{\label{FigNh}The occupation numbers, $n^d$ and $n^f$, vs doping at finite bandwidth of $f$-electrons. Solid (dashed) curves correspond to the DMFT (Hubbard-I) approximation. The parameters of the model are: $W_f/W=0.2$, $U/W=2.5$, $\epsilon_f/W=0.1$, and $T/W=0.01$. In the inset the system's free energy is shown. Solid (dashed) curve corresponds to the DMFT (Hubbard-I) approximation. The Maxwell constructions for both curves are shown by dot-dashed lines.}
\end{figure}

In this section I consider a more general model where $f$-electrons have a finite bandwidth. The model Hamiltonian is the sum of the Falicov-Kimball model's Hamiltonian~\eqref{HFK}, and the term describing the kinetic energy of the $f$-electrons:
\begin{equation}\label{HH}
H_f=-\frac{W_f}{12}\sum_{\langle ij\rangle}\left(f^{\dag}_{i}f_{j}+H.c.\right),
\end{equation}
where $W_f$ is the bandwidth of free $f$-electrons. The $\mathbf{k}$-dependence of the $f$-electrons spectrum is assumed to be the same as for $d$-electrons. Such a model can be considered as a two-band Hubbard model for spinless fermions.

In the Hubbard-I approximation, the expression for $G_d(i\omega_n,\mathbf{k})$ remains the same, but the $f$-electrons Greens's function, $G_f$, and the dispersion (for interacted electrons) take the form of~\eqref{GdHI} and~\eqref{HIspec}, respectively, with the replacement $n^f\to n^d$, $\mu\to\mu+\epsilon_f$ and $\varepsilon(\mathbf{k})\to\varepsilon(\mathbf{k})W_f/W$.

The changes in the DMFT scheme are more significant. Now one needs to introduce two dynamical mean-fields, $\lambda_d(i\omega_n)$ and $\lambda_f(i\omega_n)$. These fields, and the local Greens's functions, $D_d(i\omega_n)$ and $D_f(i\omega_n)$ [together with self-energies $\Sigma_d(i\omega_n)$ and $\Sigma_f(i\omega_n)$], are calculated numerically by the `exact diagonalization' method~\cite{DMFTreview,EDprl}. In the iteration scheme, the total number of electrons is fixed, and for each iteration the chemical potential is recalculated from given $\lambda_{d,f}(i\omega_n)$ and $\Sigma_{d,f}(i\omega_n)$.

The detailed comparison of two approximations for such this model is beyond the scope of the present paper. Here, I restrict myself by the analysis of the stability of the system toward the phase separation. The mean values $n^f$ and $n^d$, and the system's free energy $F$ as functions of doping can be calculated in the way similar to that described in Sections~\ref{SecHI} and~\ref{SecDMFT}. The typical curves are presented in figure~\ref{FigNh}. The value of the $f$-electrons band is taken to be $W_f=0.2W$, and the energy $\epsilon_f$($=0.1W$) characterizing the relative position of the bands corresponds to figure~\ref{FigN}. Both approximations give qualitatively the same behavior of occupation numbers and the free energy, very similar to that presented in figures~\ref{FigN} and~\ref{FigF}. The quantitative difference can be explained again by the renormalization of the energy $\epsilon_f$ appeared in the DMFT approximation and lost in the Hubbard-I one. In both approximations the system turns out to be unstable toward separation into the phases with $n^f\approx0$ and $n^d\approx0$. The region of phase separation depends on the value of $W_f$: it narrows when $W_f$ increases and disappears at some critical value $W_f^{*}$ (which depends on other model parameters). The Hubbard-I approximation gives $W_f^{*}/W\approx0.3$ for the model parameters corresponding to figure~\ref{FigNh}, while the DMFT approximation gives smaller value, $W_f^{*}/W\approx0.25$.

\section{Conclusions}

In conclusion, I compared the results giving by the Hubbard-I and DMFT approximations for the spinless Falicov-Kimball model in the simple cubic lattice. The itinerant Green's function, the evolution of the system with doping, and the possibility of the phase separation were considered. It was shown that both approximations coincide in the narrow band limit, $W\to0$. For the finite $W$ and the strong coupling, $U\gg W$, both approximations do not contradict to each other giving qualitatively similar results in the spectral properties of itinerant electrons and the behavior of the system with doping. Both approximations predict the instability of the homogeneous state toward the phase separation in the wide range of model parameters. It was also shown that the system remains unstable toward the phase separation even for finite bandwidth $W_f$ of heavy electrons. This indicates that the phase separation phenomenon is the inherent feature of Hubbard-like models with wide and narrow bands, and not an artefact of this or that approximation scheme.



Author would like to thank A.V. Rozhkov, A.L.  Rakhmanov, and K.I. Kugel for stimulating discussions. This work was supported by the Russian Foundation for Basic Research (grants Nos.~11-02-00708, 11-02-91335, 11-02-00741, 12-02-92100, and 12-02-31400) and the Dynasty Foundation.


\end{document}